\documentclass[twoside]{article}
\usepackage{fleqn,espcrc2,epsfig}
\usepackage{graphicx}
\usepackage[figuresright]{rotating}

\newcommand{\AmS}{{\protect\the\textfont2
  A\kern-.1667em\lower.5ex\hbox{M}\kern-.125emS}}
\def\slash#1{#1 \hskip-0.45em /}

\def\beq{\begin{eqnarray}}
\def\eeq{\end{eqnarray}}

\title{Heavy-to-light form factors:\\ Symmetries at large recoil and
calculation of $\alpha_s$ corrections\thanks{Contribution
to {\sf IV.\ International Conference on \it Hyperons, Charm and 
Beauty Hadrons\/, \sf Valencia~2000}\/;\protect\newline 
[hep-ph/0008272].}}

\author{%
Th.~Feldmann\address{Institut f\"ur
    Theoretische Physik~E, RWTH Aachen, 52056 Aachen, Germany}
}
       
\begin{document}

\begin{abstract}
Recently it has been shown that in the large-recoil limit new
symmetries emerge which impose various relations on form factors
that parametrise the decay of heavy $B$ mesons into light mesons.
These symmetry relations are broken by
radiative QCD corrections. We show that these corrections are
perturbatively calculable, and present results to first order
in the strong coupling constant $\alpha_s$.
\vspace{1pc}
\end{abstract}

\maketitle

\section{Introduction}

We study matrix elements of bilinear quark currents,
that are parametrised by form factors,
encoding the (long-distance) strong interaction 
effects in exclusive, semi-leptonic or radiative $B$ decays, such as 
$B\to\pi l\nu$, $B\to K^* \gamma$ etc. They also appear as 
non-perturbative parameters in the factorisation theorem for 
non-leptonic $B$ decays in the heavy quark mass limit 
\cite{Beneke:1999br}.
The knowledge of these form factors therefore helps to determine 
the CKM coupling $|V_{ub}|$, and to predict CP violating asymmetries 
and other quantities in rare $B$ decays. 

Charles {\em et al.}\/ 
have shown that 
certain symmetries apply, 
when the momentum of the final light meson is large \cite{Charles:1998dr}. 
These symmetries  
reduce the number of independent form factors, but 
they are broken by radiative corrections. 
In this contribution, which is based on work done together
with Martin Beneke \cite{Beneke:2000xx},
we give a brief derivation of the
symmetry relations and then discuss the computation of
symmetry-breaking corrections
at first order in the strong coupling constant $\alpha_s$.

\section{Large-recoil symmetries}

Let us concentrate on
the form factors for $\bar{B}$ decays into a pseudoscalar meson.
They are 
defined by the following Lorentz decompositions of the matrix 
elements:
\begin{eqnarray} 
 && \hskip-1.9em
 \langle P(p')|\bar q \, \gamma^\mu b |\bar{B}(p)  \rangle = 
\nonumber \\
&& \hskip-1.9em
 \quad 
 f_+(q^2)\left[p^\mu+p^{\prime\,\mu}-\frac{M^2\!-\!m_P^2}{q^2}\,q^\mu\right]
+ {}
\nonumber \\
&& \hskip-1.9em 
\quad  f_0(q^2)\,\frac{M^2\!-\!m_P^2}{q^2}\,q^\mu \ ,
\label{fpseudo1}
\end{eqnarray}
\begin{eqnarray}
&& \hskip-1.9em
\langle P(p')|\bar q \, \sigma^{\mu\nu} q_\nu b|\bar{B}(p)
 \rangle = \nonumber \\
&& \hskip-1.9em
\quad \frac{i f_T(q^2)}{M\!+\!m_P}\left[q^2(p^\mu+p^{\prime\,\mu})-
(M^2\!-\!m_P^2)\,q^\mu\right] \ , 
\label{fpseudo2}
\end{eqnarray}
where $M$ is the $B$ meson mass, $m_P$ the mass of the pseudoscalar 
meson and $q=p-p'$. 
%

It is useful to recapitulate the 
implications of heavy quark symmetry, when the final meson $P$ is 
also heavy, for example a $D$ meson.

\subsection{Heavy-heavy decays}

\begin{figure}[tbhp]
\vskip-0.3em
\begin{center}
(a) \hfill (b)\\
\psfig{file=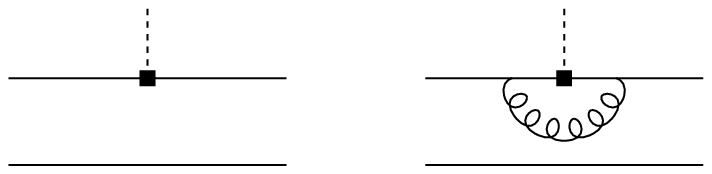, width=6cm,bb=185 630 395 715}\\[-1.5em]
\psfig{file=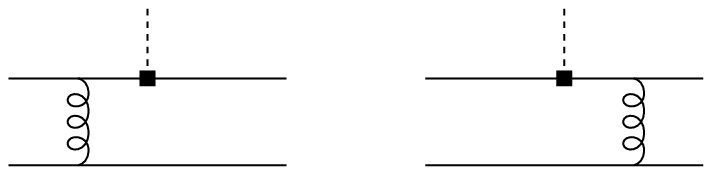, width=6cm,bb=185 670 395 755} \\[-1.5em]
(c) \hfill (d)
\end{center}
\vskip-1.8em
\caption[dummy]{\label{fig1} Different contributions to the 
$B\to P(V)$ transition. (a) Soft contribution (soft interactions with 
the spectator antiquark are not drawn). (b) Hard vertex 
renormalisation. (c,d) Hard spectator interaction.}
\end{figure}

As long as the velocity transfer to the $D$ meson remains of order 
1, we may assume that the heavy quarks interact with the 
spectator quark (and other soft degrees of freedom) exclusively via  
soft gluon 
exchanges characterised by momentum transfers much smaller than 
the heavy quark masses. Any hard interaction would imply large momentum 
of the spectator quark in the $B$ meson or $D$ meson or both, and such 
a configuration is assumed to be highly improbable. The simplifications 
that occur when heavy quarks interact only with soft gluons are 
formalised as heavy quark effective theory (HQET)
(see e.g.\ Ref.~\cite{Neubert:1994mb} and references therein) 
which implies the well-known spin and heavy flavour symmetries 
in the infinite quark mass limit 
\cite{Isgur:1989vq}.
A consequence of these symmetries is that
all form factors are related to a single function of velocity transfer, 
$\xi(v\cdot v')$, whose absolute normalisation is known at zero recoil 
($\xi(1)=1$). The heavy quark symmetries 
are violated by radiative corrections (as well as higher dimension 
operators in the effective Lagrangian), such as the one shown in 
Fig.~\ref{fig1}b. 
The symmetry breaking effects are caused only by the short-distance 
part of Fig.~\ref{fig1}b. They are accounted for by multiplicatively 
renormalising the heavy quark current in HQET.

\vskip-0.3em

In the HQET formalism the {\em hard}\/ contributions from
Fig.~\ref{fig1}c,d are assumed to be suppressed
exponentially by the $B$ and $D$ meson wave functions. 
As we shall see, 
a non-vanishing contribution from ``hard spectator interactions'' is 
one of the main differences between heavy-light  and 
heavy-heavy form factors at large recoil. 


\subsection{Heavy-light decays}

The recoil energy of the light meson, $E=(M^2-q^2)/(2 M)$, 
is of order $M/2$ if the momentum transfer is small,
$q^2 \ll M^2$ (here and in the following we have approximated
$p'^2=m_P^2 \simeq 0$). 
Let us consider the large recoil limit and assume that
the $b$ quark and the energetic light quark, created 
in the heavy-to-light transition, interact with the spectator quark (and 
other soft degrees of freedom) exclusively via soft gluon exchanges. We may 
then continue to exploit 
the heavy quark symmetries for the $b$ quark. A similar 
simplification occurs for the light quark.
To leading order in 
$\Lambda_{\rm QCD}/E$, the interaction of light 
energetic quarks with soft
gluons can be described by an effective Lagrangian
\cite{Dugan:1991de,Charles:1998dr}, 
\begin{equation}
  {\mathcal L}_{\rm eff} = \bar q_n \, \frac{\slash n_+}{2} \,
   (i \, n_- \!\cdot D) \, q_n + \mathcal{O}(1/E) \ ,
\label{leet}
\end{equation}
where $n_\pm^\mu$ are two light-like vectors ($n_+\cdot n_-=2$), and 
$q_n(x) = e^{i E ( n_- \cdot x)} \, \frac{\slash{n}_- \slash {n}_+}{4}
\,q(x)$.
In the limit $E \to \infty$ the Lagrangian~(\ref{leet}) 
leads to new symmetries that, combined with the heavy-quark symmetries,
imply non-trivial relations between the {\em soft}\/ 
contributions to the form factors \cite{Charles:1998dr}.
Using techniques familiar from HQET, we find
\begin{eqnarray}
&& \hskip-1.9em  \langle P
(p')| \bar{q}_n \, \Gamma \, b_v| \bar{B}(p)\rangle
  =
{\rm tr} \left[A_P
(E)\, \overline {\mathcal M}_{\rm L} 
    \, \Gamma \, {\mathcal M}_{\rm B} \right]
\label{trace-form} 
\end{eqnarray}
where 
\beq
 \overline{\mathcal M}_{\rm P
} &=&
        (-\gamma_5) 
\, \,\frac{\slash n_+ \slash n_-}{4}
,
\nonumber \\[0.2em]
{\mathcal M}_{\rm B} &=&  \frac{1+\slash v}{2} \,(-\gamma_5) \ ,
\label{IW-pi}
\eeq
with 
$p^\mu \simeq M v^\mu$, $p'{}^\mu \simeq E n_-^\mu$.
The most general form of the function $A_P
(E)$ is 
\begin{eqnarray}
A_P(E) &=& 2 E \,\xi_P(E), 
\end{eqnarray}
with a conveniently chosen overall normalisation. 
It follows that 
the three pseudoscalar meson form factors are all related to a single
function $\xi_P(E)$. (Analogously, the seven form factors for decays
into light vector mesons can be expressed in terms of only
two unknown functions, $\xi_\perp(E)$ and
$\xi_\parallel(E)$, see also Ref.~\cite{Charles:1998dr}.) 
Performing 
the trace in Eq.~(\ref{trace-form}), and comparing with the form
factor definition~(\ref{fpseudo1},\ref{fpseudo2}), we obtain,
for instance, 
\beq
&& \hskip-1.9em
f_+(q^2) = \frac{M}{2E} \, f_0(q^2) = \frac{M}{M+m_P} \, f_T(q^2) = 
\xi_P(E) 
\cr &&
\label{pirelation}
\eeq

There is an important distinction between the
effective Lagrangian for heavy quarks and for energetic light 
quarks. A configuration where one light quark carries most
of the energy is atypical. Therefore the symmetries of the
interaction are not realised in the hadronic spectrum. 
Furthermore, 
the probability that such an asymmetric parton configuration
hadronises into a light meson depends on the energy of the meson.
Hence, the soft contributions to the form factors are energy-dependent
functions, whose absolute normalisation is not known.
This is to be 
contrasted to the case of heavy-heavy form factors, for which 
the spin symmetry relates 
pseudoscalar and vector mesons, and 
the Isgur-Wise form factor $\xi(v\cdot v')$ is independent of the 
heavy quark mass.
The scaling law for the soft contribution can be derived in different 
ways, but all of them make use of the endpoint behaviour of the 
pion's light-cone distribution amplitude \cite{Lepage:1980fj}. If one 
{\em assumes}\/ that the distribution amplitude vanishes linearly 
with the longitudinal momentum fraction of the spectator quark, 
as is suggested by its asymptotic form, one finds
\begin{equation}
\label{softbpi}
\xi_P(E) 
\sim \frac{M^{1/2} \, \Lambda_{\rm QCD}^{3/2}}{E^2} \ .
\end{equation}

As in the case of a heavy-to-heavy transition, there will be a vertex 
correction to Eq.~(\ref{pirelation})
(of the type shown at one loop in Fig.~\ref{fig1}b). The 
hard part of these diagrams does not respect the symmetry relations, 
but it can be accounted for in perturbation theory by multiplicatively
renormalising the heavy-to-light current 
just as in the case of a heavy-to-heavy transition.

The important new element of the discussion is provided by hard
spectator interaction in Figs.~\ref{fig1}c and \ref{fig1}d. 
This allows the light meson to be formed in a preferred
configuration, in which the momentum is distributed nearly equally
between the two quarks. 
Since both quarks 
have momentum of order $M$,
and the gluon in 
Figs.~\ref{fig1}c and \ref{fig1}d has 
virtuality of order $(M\Lambda_{\rm QCD})$, this contribution can be 
computed within the hard-scattering approach to exclusive processes 
\cite{Lepage:1980fj,Efremov:1980qk,Chernyak:1984eg}. 
The resulting scaling behaviour for the pseudoscalar meson 
form factors is (see, for instance, Refs.~\cite{Chernyak:1984eg,Dahm:1995ne})
\begin{equation}
\label{hardbpi}
f_{i \,\rm hard}(q^2 \simeq 0) 
\sim \alpha_s(\sqrt{ M\Lambda_{\rm QCD}})\,
\Big(\frac{\Lambda_{\rm QCD}}{M}\Big)^{\!3/2} .
\end{equation}
We can summarise this discussion by the 
following factorisation formula for a heavy-light 
form factor at large recoil, and at leading order in $1/M$:
\beq
\label{fff}
f_i(q^2)  &=& C_i \, \xi_P(E)  + 
\Phi_B \otimes T_i \otimes \Phi_P,
\eeq
where $i=\{+,0,T\}$. Analogous formulas hold for the form factors that
parametrise the decays into light vector mesons \cite{Beneke:2000xx}.
Here $\xi_P(E)$ is the soft part of the form factor, to which the 
symmetries discussed above apply; $T_i$ is a hard-scattering kernel 
(with the endpoint divergence regulated in a certain manner),
convoluted with the light-cone distribution amplitudes of the 
$B$ meson and the light pseudoscalar meson; $C_i$
contains the hard vertex renormalisation. 
Eq.~(\ref{hardbpi}) implies that 
the hard spectator interaction (Figs.~\ref{fig1}c and \ref{fig1}d) 
is suppressed by one power of $\alpha_s$
relative to the soft contribution (Fig.~\ref{fig1}a). 
Hence the form factor relations like Eq.~(\ref{pirelation})  
are indeed correct at leading order 
in $1/M$ {\em and}\/ $\alpha_s$.

We are going to calculate the order $\alpha_s$ corrections
to Eq.~(\ref{pirelation}) in the full theory and identify the
soft contributions to be absorbed into $\xi_P$.
We will not explicitly match the effective theory onto QCD, but,
instead, define a factorisation scheme by
imposing the convenient condition
\beq
 && \hskip-1.9em f_+ \equiv  \xi_P \ . 
\label{constraint}
\eeq
In this way, also some hard contributions
are  absorbed into
$\xi_P$. 
In particular, we circumvent some subtleties, arising in
the effective theory defined by Eq.~(\ref{leet}), that 
have been addressed in \cite{Bauer:2000ew}.

\section{Results}

Calculating the vertex corrections from Fig.~\ref{fig1}b,
using $\overline{\rm MS}$ renormalisation,
and comparing with Eqs.~(\ref{fff},\ref{constraint}),
we obtain ($C_+ \equiv 0$)
\begin{equation}
  C_0 = \frac{2E}{M} \left(1 + \frac{\alpha_s C_F}{4\pi} 
\, ( 2- 2 \, L) \right) \ ,
\end{equation}
\begin{equation}
C_T =\frac{M+m_P}{M} \left( 
1 +  \frac{\alpha_s C_F}{4\pi} \, (\ln\frac{M^2}{\mu^2} + 2 \, L )
\right)  
\end{equation}
where we have introduced the abbreviation 
$L = -\frac{2E}{M-2E} \ln\frac{2E}{M}$.
Note that the tensor form factors depend on the
renormalisation scale $\mu$. 

The hard-scattering corrections are calculated from
Figs.~\ref{fig1}c,d by convoluting the hard-scattering amplitude
with the light-cone wave functions of heavy and light mesons.
With Eqs.~(\ref{fff},\ref{constraint}) this yields
($\Phi_B \otimes T_+ \otimes \Phi_P \equiv 0$)
\begin{equation}
 \Phi_B \otimes T_0 \otimes \Phi_P 
 =   \frac{\alpha_s C_F}{4\pi} \, \frac{q^2}{2EM} \, \Delta F_P \ ,
\end{equation}
\begin{equation}
 \Phi_B \otimes T_T \otimes \Phi_P 
 =   - \frac{\alpha_s C_F}{4\pi} \, \frac{M+m_P}{2E} \, \Delta F_P \ ,
\end{equation}
where we have defined the abbreviation 
\begin{equation}
\Delta F_P
    \equiv \frac{8 \pi^2 \, f_B f_P}{N_C \, M} 
          \, \langle l_+^{-1} \rangle_B \,
          \langle \bar u^{-1} \rangle_{P} \ .
\label{pihard}
\end{equation}
Here $l_+$ is the light-cone plus component of the
spectator quark in the $B$ meson, and
$\langle l_+^{-1} \rangle_B$ 
the corresponding moment with its light-cone
distribution amplitude.
The same moment appears in the leading-order contribution to
$B \to \ell\nu\gamma$ decays~\cite{Korchemsky:1999qb}, and
determines the leading non-factorisable corrections to 
$B\to\pi\pi$ decays \cite{Beneke:1999br}. 
The moment $\langle \bar u^{-1} \rangle_P$, 
with $\bar u$ being the 
longitudinal momentum fraction of the spectator quark
in the light meson, 
is accessible, for instance,
from the analysis of 
$P\gamma$ transition form factors,
see, for instance, the review in 
Ref.~\cite{Feldmann:2000uf}.

Analogous results are found for the form factors that
parametrise decays into light vector mesons \cite{Beneke:2000xx}.
We quote only one example that is important for the analysis
of the forward-backward asymmetry zero in
the rare decay $B \to K^*\ell^+\ell^-$.
The form factor ratios that enter
the relation between the Wilson coefficients
$C_7^{\rm eff}$ and $C_9^{\rm eff}$~\cite{Ali:1999mm,Burdman:1998mk}
receive the following $\alpha_s$ correction 
\begin{eqnarray}
&& \hskip-1.9em 
 \frac{M+m_V}{M} \, \frac{T_1}{V} = \frac{M}{M+m_V} \, \frac{T_2}{A_1}
 =
\nonumber \\
&&\hskip-1.9em \quad 1 + \frac{\alpha_s C_F}{4\pi} [ \ln \frac{M^2}{\mu^2} - L]
 + \frac{\alpha_s C_F}{4\pi} \, \frac{M}{4E} \,
\frac{\Delta F_\perp}{\xi_\perp} 
\label{rhoratio}
\end{eqnarray}
where a similar quantity as in Eq.~(\ref{pihard}),
\begin{equation}
\Delta F_\perp
    \equiv \frac{8 \pi^2 \, f_B f_{\perp}}{N_C \, M} 
         \, \langle l_+^{-1} \rangle_B \,
          \langle \bar u^{-1} \rangle_{\perp} \ ,
\label{DeltaFperp}
\end{equation}
 enters, with $  \langle \bar u^{-1} \rangle_{\perp} $ being a moment of
the distribution amplitude of a transversely polarised vector meson.
Inserting standard values for the parameters in 
Eqs.~(\ref{rhoratio},\ref{DeltaFperp}), the 
$\alpha_s$ corrections in Eq.~(\ref{rhoratio}) amount to
about 5\%.

In summary, we have shown that, in the kinematic region where
the momentum transfer to the light meson is {\em large}\/,
heavy-to-light form factors can be described by a 
factorisation formula~(\ref{fff}) which is based on the 
symmetries that arise in the large-recoil/heavy-quark limit. 
This implies that $\alpha_s$
corrections to symmetry relations 
are calculable in a systematic way: 
i) Vertex corrections to the heavy-to-light current can
be treated in an analogous way as in heavy quark effective
theory. ii) Hard rescattering with the spectator quark
is described by the hard-scattering approach 
which involves light-cone distribution amplitudes of the participating
mesons.
Typically, to first order in the strong
coupling constant, these corrections amount to a few percent.
Further details can be found in Ref.~\cite{Beneke:2000xx}.

\end{document}